\documentclass[letterpaper,10pt]{article} 
\usepackage{multibib}
\usepackage{opticameet3} 

\newcommand\authormark[1]{\textsuperscript{#1}}

\usepackage{amsmath,amssymb}
\usepackage[colorlinks=true,bookmarks=false,citecolor=blue,urlcolor=blue]{hyperref} 

\begin{document}

\title{Distributed $g^{(2)}$ Retrieval with Atomic Clocks: Eliminating Conventional Sync Protocols}


\author{
Md Mehdi Hassan,\authormark{1*}
Jacob E. Humberd,\authormark{2,3}
Mohmad Junaid Ul Haq,\authormark{2,3}
Noah A. Crum,\authormark{1}
George Siopsis,\authormark{1}
and Tian Li,\authormark{2,3$\dagger$} }

\address{
\authormark{1}Department of Physics \& Astronomy, University of Tennessee, Knoxville, TN 37996, USA\\
\authormark{2}Department of Physics \& Astronomy, University of Tennessee, Chattanooga, TN 37403, USA\\
\authormark{3}UTC Quantum Center, University of Tennessee, Chattanooga, TN 37403, USA\\
}

\email{
\authormark{*}mhassa11@vols.utk.edu,\;
\authormark{$\dagger$}tian-li@utc.edu
} 


\begin{abstract}
We demonstrate a method to measure coincidences between polarization-entangled photons distributed to distant locations, eliminating traditional synchronization by employing a compact, chip-scale atomic clock for precise timing.  \\\\
\end{abstract}

Distributed quantum optical measurements fundamentally require precise timing alignment. While modern time taggers achieve picosecond-level jitter for local measurements, this precision alone is insufficient in networked systems: independent local oscillators inevitably drift apart. Conventional solutions, such as GNSS/GPS and PTP-based protocols like White Rabbit\cite{JANSWEIJER2013187} can provide down to sub-nanosecond synchronization but require dedicated infrastructure and co-propagating classical signals, introducing vulnerability to crosstalk and interference. Recent advances in chip-scale Rb atomic clocks offer excellent holdover stability in a compact form factor\cite{macsa57}. Once their frequencies are tuned, they can maintain mutual coherence long enough to enable distributed quantum-correlation measurements without any explicit timing link. In this paper, we report “synctonizing” two independent time taggers using only Rb atomic clocks, with no GNSS reference or physical timing link, and recovering correlations between polarization-entangled photons separated by a 10 km fiber spool.

\begin{figure}[htbp]
\centering
\includegraphics[width=0.95\columnwidth]{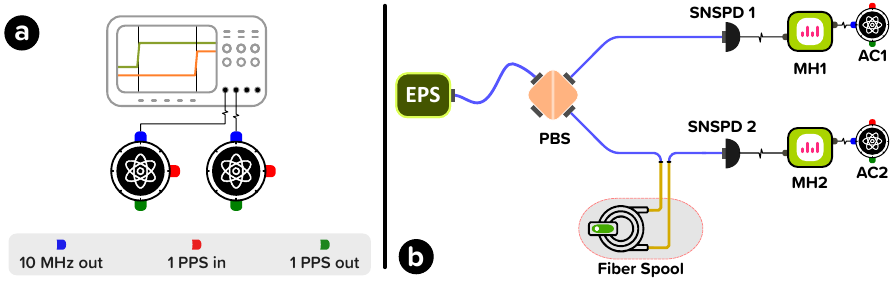}
\caption{Experimental setup. \textbf{(a)} Analysis of frequency drift when one atomic clock (right) is digitally tuned and disciplined with respect to the other (left). \textbf{(b)} The main setup, where a polarizing beam splitter (PBS) splits the correlated photons. The heralding photon is detected locally by a single-photon detector (SNSPD 1) connected to a MultiHarp 150 time tagger (MH1), disciplined by atomic clock AC1. The signal photon travels through a 10 km fiber spool before reaching SNSPD 2 and is then time-tagged via MH2, disciplined by AC2.}
  \label{fig:setup}
\end{figure}

Figure \ref{fig:setup} shows the experimental setup, which consists entirely of off‑the‑shelf, commercially available components. The Qubittek Type-II biphoton source (EPS) generates $\sim 3\times10^6$ degenerate, correlated photon pairs per second \cite{reaz2025} at 1570 nm. The photon pairs are separated using a polarizing beam splitter: the herald photons are detected locally by a single-photon detector, while the signal photons are sent through a 10 km fiber spool to a second detector that emulates a remote quantum node. Time tagging is performed using a MultiHarp 150 unit with 5 ns timing resolution. By default, the time bins are referenced to the internal oscillator of the device; however, several timing-reference options are available, including an external reference of 10 MHz, a GNSS 1PPS signal, and the White Rabbit operation (either master or slave). In our experiment, the 10 MHz signal from an atomic clock was used as the reference. There are three principal methods for calibrating the frequency and phase of the rubidium oscillators: (1) using a GPS 1PPS signal, (2) digital tuning, and (3) analog tuning. In this work, we employed digital tuning with the aid of an oscilloscope, in order to avoid relying on the GPS signal and to achieve faster disciplining. Once the clocks were digitally tuned to eliminate visible frequency drift, they were monitored for an additional hour to quantify the residual drift rate and compute the Allan deviation. \\

\begin{figure}[htbp]
  \centering
  \includegraphics[width=1\columnwidth]{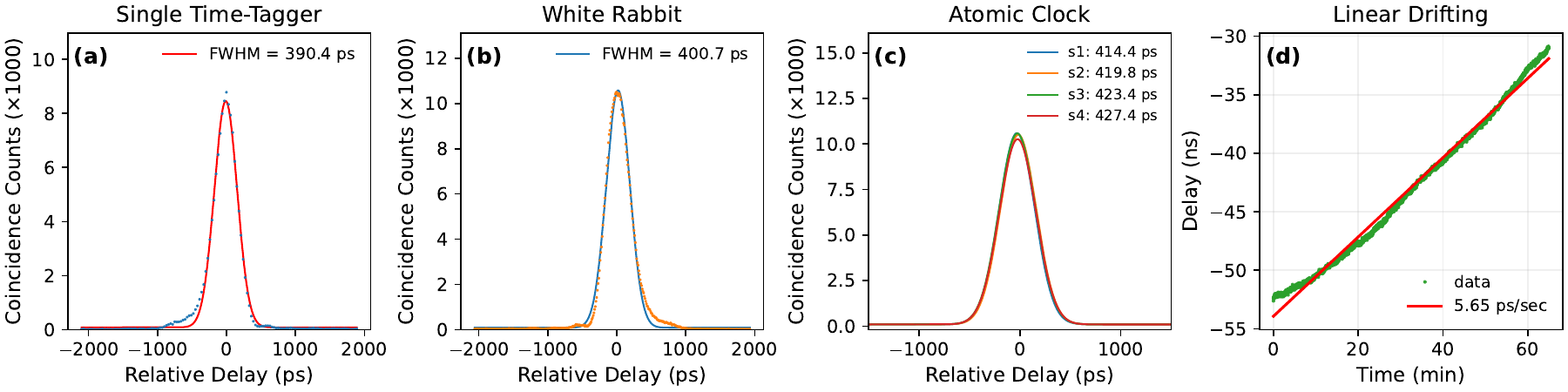}
  \caption{$g^{(2)}$ correlation analysis: 
  (a) Both SNSPDs are connected to a single MultiHarp 150 time-tagger; 
  (b) the SNSPDs are connected to separate, distant MultiHarps, disciplined by a White Rabbit switch; 
  (c) the SNSPDs are connected to separate, distant MultiHarps, each disciplined by Rb oscillators; 
  (d) linear drift statistics of the relative frequency.}
  \label{fig:analysis}
\end{figure}

Figure \ref{fig:analysis} summarizes the $g^{(2)}$ correlation results. As expected, when both detectors share a single time-tagger, the histogram in Fig.~\ref{fig:analysis}(a) exhibits the narrowest FWHM. The White Rabbit--disciplined configuration in Fig.~\ref{fig:analysis}(b) shows slightly broader correlations due to network-induced residual timing offsets but remains superior to the Rb-disciplined case. In Fig.~\ref{fig:analysis}(c), the drifting of independent Rb oscillators manifests as a gradual FWHM increase of 13~ps over $\sim1$ hour. Here, we recorded 5-second-long data sets during this period and selected four samples (s$_1$--s$_4$, with s$_1$ being the earliest) roughly 15 minutes apart. This behavior is in agreement with the independently measured drift rate of 5.65 ps/s: a small frequency offset between the oscillators broadens the correlation peak, and a sufficiently large offset ultimately washes out the observable coincidence signal.

Using GNSS signals can improve long-term stability and make the data analysis substantially easier. However, our approach demonstrates that short-duration quantum-correlation measurements can be achieved with only pairwise-tuned, chip-scale atomic clocks, eliminating external synchronization links and reducing vulnerability to signal-jamming attacks relevant to secure QKD. 

\section*{Acknowledgments}

JEH and TL acknowledge support from the U.S. National Science Foundation (NSF) through the ExpandQISE program under Award No. 2426699, and CCSS program under Award No. 2503630. MJUH and TL also acknowledge support from the U.S. National Institute of Standards and Technology (NIST) through the CIPP program under Award No. 60NANB24D218. NAC, MMH, and GS acknowledge support from the U.S. Department of Energy, Office of Science, Office of Advanced Scientific Computing Research, through the Quantum Internet to Accelerate Scientific Discovery Program under Field Work Proposal 3ERKJ381. NAC, MMH, and GS also acknowledge support from the NSF under Award No. DGE-2152168.

\bibliography{references}

\end{document}